\documentstyle[fleqn,twoside]{article}
\def\be{\begin{equation}}
\def\ee{\end{equation}}
\def\bi{\bibitem}
\begin{document}
\title{Quantum Mechanical Formulation Of Quantum Cosmology For 
Brane-World Effective Action} 
\author{Abhik Kumar Sanyal} 
\maketitle
\noindent
\begin{center}
Dept. of Physics, Jangipur College, Murshidabad,
\noindent
West Bengal, India - 742213. \\
\noindent
and\\
\noindent
Relativity and Cosmology Research Centre\\ 
\noindent
Department of Physics, Jadavpur University \\
\noindent
Kolkata - 700 032, India\\
\noindent
e-mail:  aks@juphys.ernet.in, and sanyal\_ ak@yahoo.com\\
\end{center}
\noindent
\begin{abstract}
Canonical quantization of the Brane-World effective action presented by 
Kanno and Soda containing higher order curvature invariant terms, has been 
performed. It requires introduction of an auxiliary variable. As observed 
in a series of publications by Sanyal and Modak, here again we infer that 
properly chosen auxiliary variable leads to a Schr$\ddot{o}$dinger like 
equation where the kinetic part of a canonical variable disentangles from 
the rest of the variables giving a natural quantum mechanical flavour of 
time. Further, the effective Hamiltonian turns out to be hermitian, leading 
to the continuity equation. Thus, a quantum mechanical probability 
interpretation is plausible. Finally, the extremization of the effective 
potential leads to Einstein's equation and a well behaved classical solution, 
which is a desirable feature of the gravitational action containing higher 
order curvature invariant terms. 
\\ 
\noindent
{PACS NOS.04.50+h,04.20.Ha,98.80.Hw}
\end{abstract}

\section{\bf{Introduction}}

The aim of quantum cosmology is to give a quantum mechanical meaning to 
the existing classical cosmological models.  Despite intense research in 
this field all the attempts in this regard went in vain. In a series of 
recent publications, \cite{a:p}, \cite{a:c} and \cite{a:pl}, it has been 
shown that, there still exists the possibility of giving a quantum 
mechanical interpretation to quantum cosmology at least in an isotropic 
and homogeneous minisuperspace model.  
\par
Under a $(3+1)$ decomposition it is always possible to express the 
space-time metric locally in the form,
\be
ds^2=-(n^2-n^{i}n_{i})dt^2+2n_{i}dx^{i~2}+h_{ij}dx^{i}dx^{j}.
\ee
where $n$ and $n_{i}$ are the lapse function and the shift vector 
respectively and $h_{ij}$ is the induced metric on the submanifold. 
In view of (1) the Einstein-Hilbert action
\be
S=\frac{1}{16\pi G}\int~\sqrt{-g}~^4{R}~ d^4{x} -\frac{1}{8\pi 
G}\int~\sqrt{h}~k~d^3{x}
\ee
leads to Hamiltonian constraint equation $H = 0$ rather than the
Hamiltonian. In action (2) $k$ is the trace of the extrinsic curvature 
and $h$ is the determinant of the metric on the submanifold. The second 
term on the right hand side is the well-known Gibbons-Hawking term.The 
canonical quantization of the Hamiltonian constraint equation yields zero 
energy Schr$\ddot{o}$dinger equation, known as the Wheeler-deWitt equation.
\be
-[\hbar ^2 G_{ijkl}\frac{\partial}{\partial 
h_{ij}}\frac{\partial}{\partial h_{kl}}+\sqrt{h}~^3 R]\Psi(^3 G)=0
\ee
where, $G_{ijkl}$ is the infinite dimensional metric of the so called 
superspace. Since classical cosmological models are described by 
minisuperspace models, therefore, the complication arising out of infinite 
dimensional superspace can be avoided by considering $h_{ij}$'s to be 
functions of time alone, reducing the problem to the finite dimensional 
minisuperspace. Effectively, it reduces the problem of quantum cosmology to 
that of quantum mechanics that can be solved with appropriate boundary 
condition. Such a boundary condition was proposed by Hartle-Hawking 
\cite{hh:p} by going over to the euclidean functional integral.
\par
Now, the Euclidean functional integral for the Einstein-Hilbert action (2) is 
not bounded from below, as such it diverges badly and the programme of 
reducing the problem of quantum cosmology to that of quantum mechanics, where 
one can construct the Hilbert space and give a probabilistic interpretation 
fails. This is one of the reasons to modify action (2) by the introduction
of higher order curvature invariant terms in a way such that the Euclidean 
functional integral converges. Horowitz \cite{ho:p} has shown that 
introduction of $R^2$ term in the action leads to Schr$\ddot o$dinger like 
equation in the sense that one of the parameters of the theory acts as time 
parameter. Stell \cite{k:p} claimed that the action $\int d^4 x~\sqrt{-g} 
[A C_{ijkl}^2+B~ ^4{R}+C~ ^4{R}^2]$ is perturbatively renormalizable in 
$4$-dimensions. Starobinsky \cite{s:pl} considered Einstein's equation with 
quantum one loop correction which contains $R^2$ term and obtained 
Inflationary solution. Hawking and Luttrell \cite{hl:n} have shown that $R^2$ 
action under a conformal transformation leads to Einstein-Hilbert action 
minimally coupled with a Scalar field. Recently, the importance of considering 
$R^2$ term in the action has further been increased as it has been observed 
that the $4$-dimensional Brane world effective action contains such term. 
\par 
The most attractive feature of higher order theory of gravity has been 
demonstrated in our recent works \cite{a:p},\cite{a:c} and \cite{a:pl}. 
The Hamiltonian formulation of the action containing higher order curvature 
invariant terms requires the introduction of an auxiliary variable. In terms 
of this variable, as the action is expressed in canonical form, one of the true
degrees of freedom disentangles from the kinetic part, giving rise to a 
quantum mechanical flavour of time. However, in \cite{a:p}, it has been 
shown 
that one can even introduce auxiliary variable in Einstein-Hilbert action 
without Gibbons-Hawking term. The classical field equations remain unchanged 
as expected, since total derivative terms in no way affect them. However, it 
leads to totally wrong Wheeler-deWitt equation. Now, it is not possible 
to find a general Gibbons-Hawking type surface term like one above, for 
gravitational actions containing higher order curvature invariant terms. As 
a consequence, one can introduce the auxiliary variable straight into the 
action giving birth to wrong quantum dynamics. This result has been 
illustrated earlier \cite{a:p},\cite{a:c} and \cite{a:pl}. Total derivative 
terms thus play important role in quantum physics and so we conclude 
that all such removable terms appearing in the minisuperspace model under 
consideration should be removed prior to the introduction of auxiliary 
variable to get the correct and unique quantum description of the theory. 
This important fact has not been taken care of by the earlier 
workers \cite{hl:n}, \cite{ho:p} in this field. This proposal when taken 
up for quantization, yields Schr$\ddot{o}$dinger like Wheeler-deWitt 
equation with excellant features. The effective Hamiltonian turns out to be 
hermitian and the continuity equation now identifies the nature of space and 
time like variables in the Robertson-Walker minisuperspace model under 
consideration. This establishes a quantum mechanical idea of probability and 
current densities in quantum cosmology. Further, an effective potential 
emerges in the process, whose extremization yields vacuum Einstein's 
equation, which is ofcourse a desirable feature in the weak energy limit 
of higher order gravity theory. Thus we conclude that the quantum mechanical 
formulation of quantum cosmology requires the modification of the 
Einstein-Hilbert action by introducing higher order curvature invariant terms
at least in the Robertson-Walker minisuperspace model.
\par
In view of such excellent results it seems interesting to take up the proposal
to quantize the $4$-dimensional brane world effective action \cite{k:s} which 
contains higher order curvature invariant terms as well. 
In the last few years there has been lot of investigation on the brane world 
scenario particularly due to an interesting model proposed by  Randall and 
Sundrum \cite{r:s} based on the following action,
\be
S=\frac{1}{16\pi G_{N}l}\int~d^5 x(^5 R+\frac{12}{l^2})-\sigma\int~d^4 x \sqrt{-h}
+\int~d^4 x \sqrt{-h}L_{matter}.
\ee
which dictates that gravity can propagate everywhere in all the dimensions 
while the standard model matter is restricted in the $4$-dimensional space time
only called brane. In the above action the $4$-dimensional brane with tension 
$\sigma$ is assumed to be embedded in the $5$-dimensional assymptotically anti 
de-Sitter bulk with a curvature scale $l$. However, for the understanding 
of a lot of fundamental problems we need only the low energy $4$-dimensional
effective action, which is expected to contain higher order curvature invariant 
terms. There are different avenues to find such action, which is not our 
present deal. However, in a recent interesting paper Kanno and Soda 
\cite{k:s} have derived the following effective action by the use of low 
energy iteration scheme,
\be   
S_{e}=\frac{1}{16\pi G_{N}}\int d^4 x\sqrt{-g}[R+\frac{\beta 
l^2}{3}R^2+(\delta-1/4)l^2(R_{\mu\nu}R^{\mu\nu}-\frac{R^2}{3})] 
+ S_{matter}+ S_{cft}, 
\ee
where, $S_{cft}$ is the action corresponding to some appropriate conformal 
field theory. Our aim is to canonically quantize this $4$-dimensional 
effective action (5). Now one can raise the very fundamental question of 
the viability of quantising an already effective action. Here we present a 
favourable argument. The above effective action (5) reduces to the 
Einstein-Hilbert action being coupled to some matter field only in even 
lower energy limit. We know that our present universe is well explained by 
the standard model. So, such an effective action has nothing to do with 
the present day universe. Further, the low energy iteration scheme that 
has been carried out to find the above action (5) may not be low enough to 
turn the visible $4$-dimensional world classical. Rather, such an effective 
action (5) can play some important roles only in the quantum era of the 
$4$ dimensional world and one might obtain the standard model results 
through semiclassical approximation.
\par
The following section has been devoted in quantising the above action (5) 
and to extremize the effective potential to present some standard classical 
solutions which are desirable features of higher order gravity theory. 
In section (3) we present the continuity equation to give the quantum 
mechanical probability interpretation. A brief summary of our work has
been presented in section (4).   

\section{\bf{Wheeler-deWitt equation and extremization of the effective 
potential}}
We consider the 4-dimensional effective Brane-World action (5) given by 
Kanno and Soda \cite{k:s}, in the following form,  
\be
S= \frac{1}{16\pi G_{N}}\int d^4 x\sqrt{-h}
[R-3\phi_{,\mu}\phi^{,\mu}-3e^{2\phi}\chi_{,\mu}\chi^{,\mu}
+\frac{\beta l^2}{3}R^2+(\delta-\frac{1}{4}l^2)(R_{\mu\nu}R^{\mu\nu}
-\frac{1}{3}R^2)],
\ee
where, in the matter sector we have taken axion($\chi$)-dilaton($\phi$) 
field. In the closed Robertson-Walker metric
\be
ds^2=-dt^2+a(t)^2[d\chi^2-sin^2 \chi(d\theta^2+sin^2 \theta 
d\phi^2)] 
\ee
the Ricci scalar is
\be
R=6(\frac{\ddot{a}}{a}+\frac{\dot{a}^2}{a^2}+\frac{1}{a^2})
\ee
It is interesting to note that the last term in the action (6), viz. 
$\frac{(\delta-1/4)l^2}{16\pi G_{N}}\int 
d^4 x\sqrt{-h}(R_{\mu\nu}R^{\mu\nu}-R^2 /3)$ 
does not contribute to the field 
equation, since it can be integrated out by parts to yield a 
total derivative term, viz. $\Sigma_{2} = 
-2M(\delta-1/4)l^2(\frac{1}{3}\dot{a}^3+a)$, where $M = \frac{3\pi}{4G}$. 
So we are left with the following form of the action (6), 
\be
S= \frac{1}{16\pi G_{N}}\int d^4 x\sqrt{-h}
[R-3\phi_{,\mu}\phi^{,\mu}-3e^{2\phi}\chi_{,\mu}\chi^{,\mu}
+\frac{\beta l^2}{3}R^2]+\Sigma_{2},
\ee
The quantum cosmology of this form of action, although not in the context 
of brane-world scenario, has been studied earlier by Sanyal and modak 
\cite{a:c} in the conformal form of the above metric (7). We proceed here 
to do the same work in proper time co-ordinate with a different type of 
matter field and in a different context of brane-world scenario. 
Substitution of the form of the Ricci scalar ($R$) from (8) in the above 
form of the action (9) leads to, 
\be
S=M\int [\{(a^2 
\ddot{a}+a\dot{a}^2+a+\gamma 
a\ddot{a}^2+2\gamma\dot{a}^2\ddot{a})+2\gamma 
\ddot{a}+\gamma\frac{(\dot{a}^2+1)^2}{a}+\frac{1}{2}a^3\dot{\phi}^2 
+\frac{1}{2}a^3 e^{2\phi}\dot{\chi}^2] + \Sigma_{2} , 
\ee
where, dot stands for $\frac{d}{dt}$ and $\gamma = 2\beta l^2$. 
According to our proposal, we remove all removable total derivative terms 
from the action (10), so that only lowest order terms appear in it. In the 
process we get,
\be
S=M\int 
[a-a\dot{a}^2-\gamma(a\ddot{a}^2+\frac{(\dot{a}^2+k)^2}{a})+
\frac{1}{2}(\dot{\phi}^2 +e^{2\phi}\dot{\chi}^2)a^3]+\Sigma_{1},
\ee
where, $\Sigma_{1} 
=M[\dot{a}a^2+\frac{2}{3}\gamma\dot{a}^3+2\gamma \dot{a}] + \Sigma_{2}$. 
\par
Since the action (11) can not any further be made free from the second 
derivartive terms of the field variable, so for a Hamiltonian formulation 
of such an action we need to define an auxiliary variable at this stage, 
which is
\be 
Q=-\frac{\partial{S}}{\partial{\ddot{a}}}=-2M\gamma a \ddot{a}.               
\ee
Introducing the auxiliary variable $Q$ in action (11) and expressing the 
action in the canonical form, after removing the remaining total 
derivative terms, we obtain,
\be
S=\int 
[\dot{Q}\dot{a}-\frac{Q^2}{4M\gamma a}
+M\{a-a\dot{a}^2+\gamma\frac{(\dot{a}^2+1)^2}{a}
+\frac{a^3}{2}(\dot{\phi}^2 +e^{2 \phi}\dot{chi}^2]dt+\Sigma, 
\ee
where, $\Sigma = \Sigma_{1} -\frac{Q\dot{a}}{M}$. The classical field 
equations are,
\be
Q=-2M\gamma a\ddot{a}
\ee
which is the definition of $Q$ given in (12).
\be
\ddot{\phi}+3\frac{\dot{a}}{a}\dot{\phi}-e^{2\phi}\dot{\chi}^2=0
\ee
\be
\dot{\chi}a^3 e^{2\phi}=c
\ee
where, $c$ is a constant.
\be
\ddot{Q}+\frac{M\gamma}{a^2}[(\dot{a}^2+1)(4a\ddot{a}-3\dot{a}^2+1)
+8a\dot{a}^2\ddot{a}]-\frac{Q^2}{4M\gamma a^2}
-Ma^2[2\frac{\ddot{a}}{a}+\frac{\dot{a}^2}{a^2}+\frac{1}{a^2}
+\frac{3}{2}(\dot{\phi}^2+e^{2\phi}\dot{\chi}^2)]=0
\ee
and
\be
\dot{Q}\dot{a}+\frac{Q^2}{4M\gamma a}
+\frac{M\gamma}{a}[3\dot{a}^4 +2\dot{a}^2 -1]
-Ma^3[\frac{\dot{a}^2}{a^2}+\frac{1}{a^2}-\frac{1}{2}(\dot{\phi}^2
+e^{2\phi}\dot{\chi}^2)]=0
\ee
Now equation (18) is essentially the Hamiltonian constraint equation 
which we now express in the phase space variabes as,
\be
H=P_{a}P_{Q} 
+\frac{Q^2}{4M\gamma a}-M[\gamma\frac{(P_{Q}^2+1)^2}{a}-aP_{Q}^2+a]
+\frac{1}{2Ma^3}(P_{\phi}^2+ e^{-2\phi}P_{\chi}^2)=0,
\ee
where, $P_{a}, P_{Q}$, $P_{\phi}$ and $P_{\chi}$ are the canonical 
momenta corresponding to $a, Q, \phi$ and $\chi$ respectively. 
The Wheeler-deWitt equation is obtained through the quantization of the 
Hamiltonian constraint equation (19). Note that canonical 
quantization should be performed with the basic variables, which are $a, 
\dot{a}, \phi$ and $\chi$ here.. Let us choose $\dot{a} = x$, and hence 
replace $P_{Q}$ by $x$ and $Q$ by $-P_{x}$.  It is evident that $P_{x}$ 
is the canonical momentum corresponding to $x = \dot{a}$. Hence equation 
(19) turns out to be,
\be
H=xP_{a}+\frac{P_{x}^2}{4M\gamma a}
+\frac{1}{2Ma^3}(P_{\phi}^2+e^{-2\phi}P_{\chi}^2)-M[\gamma\frac{(x^2+1)^2}{a}
-ax^2+a]. 
\ee 
As the Hamiltonian $H$ is constrained to vanish, we get the Wheeler-deWitt 
equation as,  
\be
i\hbar a\frac{\partial{\psi}}{\partial{a}}=-\frac{\hbar^2}{4M\gamma 
x}(\frac{\partial^2{\psi}}{\partial{x^2}}+
\frac{n}{x}\frac{\partial{\psi}}{\partial{x}}) -\frac{\hbar^2}{2Ma^2 
x}(\frac{\partial^2{\psi}}{\partial{\phi^2}}+
e^{-2\phi}\frac{\partial^2{\psi}}{\partial{\chi}^2})
-M[\gamma\frac{{x^2+1}^2}{x}-a^2 x+\frac{a^2}{x}]\psi.
\ee
where $n$ is the operator ordering indix. With the choice $a=e^{\alpha}$, 
the above eqn. reduces to 
\be
i\hbar\frac{\partial{\psi}}{\partial{\alpha}}=
-\frac{\hbar^2}{2M\gamma 
x}(\frac{\partial^2{\psi}}{\partial{x^2}}+
\frac{n}{x}\frac{\partial{\psi}}{\partial{x}})
-\frac{\hbar^2}{2Mx}(\frac{\partial^2{\psi}}{\partial{\phi^2}}+
e^{-2\phi}\frac{\partial^2{\psi}}{\partial{\chi}^2})e^{-2\alpha}-
M[\gamma\frac{(x^2+1)^2}{x}-\frac{(x^2-1)}{x}e^{2\alpha}]\psi.
\ee
Which essentially is of the same form for $\gamma = - \beta$ as obtained 
in our previous work \cite{a:c}. Now, one can write,
\be
i\hbar\frac{\partial{\psi}}{\partial{\alpha}}=\hat{H_{0}}\psi.
\ee
where, $\hat{H_{0}}\psi$ is given by,
\be
\hat{H_{0}}\psi=-\frac{\hbar^2}{2M\gamma x}
(\frac{\partial^2{\psi}}{\partial{x^2}}+
\frac{n}{x}\frac{\partial{\psi}}{\partial{x}})
-\frac{\hbar^2}{2Mx}(\frac{\partial^2{\psi}}{\partial{\phi^2}}+
e^{-2\phi}\frac{\partial^2{\psi}}{\partial{\chi}^2})
e^{-2\alpha}+V_{e}\psi.
\ee
The effective potential $V_{e}$ is, 
\be
V_{e}=-M[\gamma\frac{(x^2+1)^2}{x}-\frac{x^2-1}{x}e^{2\alpha}]. 
\ee
It is evident that The Wheeler-deWitt equation (23) takes the 
Schrodinger-like form, where $\alpha$ acts as the time parameter and 
$\hat{H_{0}}$ is the effective Hamiltonian given by equation (24).
This feature is also consistent with the intrinsic concept of general 
theory of relativity, as time there has no independent existence 
from geometry in describing gravitation, rather it is inbuilt in the 
theory, unlike situations encountered in the conventional classical and 
quantum mechanics, where time is an external parameter. 
\par
The effective potential can be extremized at the energy scale where
potential energy dominates over the kinetic energy, to obtain the 
following equation
\be
(x^2+1)[\gamma(3x^2-1)-e^{2\alpha}]=0.
\ee 
The above equation yields either vacuum Einstein's equation 
\be
x^2+1 = 0,
\ee
or
\be
\gamma(3x^2-1)-e^{2\alpha}=0
\ee
which admits a solution for $\gamma > 0$
\be
a = \sqrt{\gamma}sinh[\frac{(t-t_{0})}{\sqrt{3\gamma}}],
\ee
This is evidently a wonderful result showing that the extremum of the 
effective action gives well-behaved classical solutions. This is a 
desirable feature of higher order theory of gravity proving the merit of 
our proposal. 

\section{\bf{Probability and current density}}

One of the most important feature observed in quantum mechanics is that, 
the state of a system is described by a wavevector belonging to an 
abstract Hilbert space and the norm of the wavevector must be positive 
definite or zero. This idea emerged from the interpretation of the 
probability density to describe a given state from the continuity equation 
which is obtained by using the Schr$\ddot{o}$dinger equation. Probability 
interpretation follows from the simple mathematical appearence of the 
Schr$\ddot{o}$dinger equation. No such interpretation of the probability 
density in general exists in quantum cosmology, when the action contains 
terms linear in the Ricci scalar coupled with some matter field. This is 
due to the fact that there is no time a priori in the Wheeler-deWitt 
equation in a gravitational theory described by Einstein-Hilbert action. 
\par
It is to be noted that the continuity 
equation along with the conventional notion of the probability density can 
only be introduced with the proper choice of the canonical variables in such 
a way that the Hamiltonian constraint is quadratic in the canonical momenta 
along with a term linear in momentum whose canonical coordinate acts as an 
intrinsic time variable. This type of canonical quantization is possible only 
when the action contains atleast a quadratic curvature term. It turns out 
that at least in the homogeneous and isotropic minisuperspace
quantum cosmological model, $R^2$ has a generic feature to yield a 
modified Wheeler-deWitt equation that looks like usual time-dependent 
Schr$\ddot{o}$dinger equation, where an intrinsic geometric variable 
$\alpha$, related to the scale factor plays the role of time. This further 
gives rise to a quantum mechanical probability interpretation of quantum 
cosmology, as we shall show in the following.
\par 
It is to be noted that $\hat{H_{0}}$ operator given by equation (24) is hermitian 
and as a consequence we obtain the continuity equation, viz.
\be
\frac{\partial{\rho}}{\partial{\alpha}}+\bf{\nabla}\cdot{\bf{J}}=0,
\ee
where $\rho$ and $\bf{J}$ are the probability and the current densities 
respectively for the choice of the operator ordering index $n = -1$. 
$\rho$ and $\bf{J}$ are given by $\rho = \psi^*\psi$ and ${\bf{J}} = (J_{x}, 
J_{\phi}, J_{\chi})$, where
\[
J_{x}=\frac{i\hbar}{4M\gamma x}(\psi^*\psi,_{x}-\psi\psi,_{x}^*),~~~
J_{\phi}=\frac{i\hbar 
e^{-2\alpha}}{2Mx}(\psi^*\psi,_{\phi}-\psi\psi,_{\phi}^*), 
\]
and
\be
J_{\chi}=\frac{i\hbar 
e^{-2(\alpha+\phi)}}{2Mx}(\psi^*\psi,_{\chi}-\psi\psi,_{\chi}^*), 
\ee
with,
\be 
{\bf{\nabla}}=(\frac{\partial}{\partial{x}},
\frac{\partial}{\partial{\phi}},
\frac{\partial}{\partial{\chi}},).
\ee
One can also find the continuity equation for other values of the 
operator ordering index but with respect to a new variable $y$ which is 
functionally related to $x$ only. Since the above probability 
and the current densities are dynamical quantities, therefore the 
wavefunction and its derivatives should remain finite at all epoch of the 
evolution of the Universe, if and only if there are no singularities in 
the domain of quantum cosmology.
\par
Further in analogy with quantum mechanics it is noted that the 
variable $\alpha$ in equation can be identified as the time 
parameter in quantum cosmology. The variables $x 
(=\dot{\alpha}e^{\alpha}),\phi$ and $\chi$ act as spatial coordinate 
variables in this context. 
\section{Conclusion}
Quantum cosmology for an action containing $R+R^2$ term has been 
performed and published earlier \cite{a:c}. It is noted that the effective 
action for Brane-World-Cosmology contains similar geometric terms, apart from 
$(R_{\mu\nu}R^{\mu\nu}-\frac{R^2}{3})$, which does not 
contribute to the field equation in the Robertson-Walker minisuperspace 
model. Previously, the work containing $R+R^2$ term was 
performed in the conformal time coordinate. Here we perform it for the 
$4$-dimensional brane-world effective action coupled to axion-dilaton field
in the proper time co-ordinate. The most important feature of the present 
work is that a possibility has been explored to give a quantum mechanical 
interpretation of a minisuperspace cosmological model. This came out due 
to the fact that an internal time parameter has automatically been picked 
up to express the Wheeler-deWitt equation in the standard form of 
time dependent Schr$\ddot{o}$dinger equation with an effective Hamiltonian 
that is hermitian. Hence the continuity equation can be written and the 
probabilistic interpretation follows automatically. Further, the extremum 
of the effective potential produces some standard classical solutions, 
which are no less important. We observe that the $4$-dimensional effective 
brane-world action which is obtained through low energy iteration scheme, 
does in no way leads to the standard model, if the action is treated 
classically. We conclude that the effective action still remains in the  
quantum domain of the $4$-dimensional world.

\end{document}